\documentclass[numberedappendix]{emulateapj}
\usepackage{apjfonts}
\usepackage{epsfig}

\newcommand{\be}{\begin{eqnarray}}
\newcommand{\ee}{\end{eqnarray}}
\newcommand{\qc}{Q_{\rm crit}}
\newcommand{\de}{\eta_{0.5}}
\newcommand{\mdot}{\dot{M}_1}
\newcommand{\al}{\alpha_{0.1}}
\newcommand{\lp}{\left(}
\newcommand{\rp}{\right)}
\newcommand{\tc}{t_{\rm cool}}

\newcommand{\calm}{\mathcal{M}}

\newcommand{\msun}{\,M_\odot}

\newcommand{\s}{\,{\rm s}}
\newcommand{\cm}{\,{\rm cm}}
\newcommand{\g}{\,{\rm g}}

\begin{document}


\slugcomment{Accepted for publication in The Astrophysical Journal}
\shorttitle{FRAGMENTATION OF COLLAPSAR DISKS} 
\shortauthors{PIRO \& PFAHL}


\title{Fragmentation of Collapsar Disks and the Production of
Gravitational Waves}


\author{Anthony L. Piro and Eric Pfahl} 

\affil{Kavli Institute for Theoretical Physics, Kohn Hall, University
of California, Santa Barbara, CA 93106; \\ piro@kitp.ucsb.edu,
pfahl@kitp.ucsb.edu}


\begin{abstract}

We argue that gravitational instability in the outer parts of
collapsar disks may lead to fragmentation near the radius where helium
photodisintegrates, because of the strong cooling provided by this
process.  This physics sets clear physical scales for the
fragmentation conditions and the properties of gravitationally bound
clumps. Collapse of a fragment proceeds until the neutrons become
degenerate; a neutron star of mass $\approx$0.1--$1\msun$ may result.
We find that tidal disruption of a fragment and accretion by the
central black hole are too rapid to account for the durations of
observed X-ray flares from long gamma-ray bursts.  Prior to
disruption, migration of the fragment is driven by gravitational
radiation and disk viscosity, which act together to produce a unique
gravitational-wave signature.  Advanced LIGO may be able to detect
such sources within $\approx$100\,Mpc.

\end{abstract}




\section{Introduction}
\label{sec:introduction}

Observations associating long gamma-ray bursts (GRBs) with
core-collapse supernovae \citep{hjo03,sta03} provide strong support
for the collapsar scenario \citep{woo93,pac98,pwf99,mw99}.  In this
picture, the central $\approx$$3\msun$ of a massive star collapses
directly to a black hole, while infalling material with higher
specific angular momentum forms an accretion disk of mass
$\sim$$1\msun$.  Viscous stresses in the disk drive material toward
the hole at rates of $\ga$$0.1\msun\,{\rm s}^{-1}$, which
powers relativistic outflows responsible for both the GRB and the
supernova.  Although this model has many favorable generic features,
certain aspects of the stellar collapse and disk evolution remain
poorly understood or largely unexplored.  New observational results
demand augmentation of the simplest versions of the collapsar
scenario.

Recent {\em Swift} observations have revealed that X-ray flares
occurring $\approx$$10^2$--$10^4\s$ after the initial burst are quite
common \citep[e.g.,][]{obr06}.  The flares exhibit a wide range of
amplitudes and time scales, and some GRBs show multiple eruptions.
Explanations for the flares point to delayed activity of the central
engine \citep[e.g.,][]{kin05,bur05,pz06,paz06}.  In particular,
\citet{paz06} speculate that an extended collapsar disk may be
gravitationally unstable and fragment at large radii, where the
viscous time scale coincides with the onset times of the flares.  This
work prompted us to examine in more detail the physics of the
fragmentation process under the extreme conditions of collapsar disks.
We also take a fresh look at the observable consequences, with special
emphasis on gravitational-wave emission.

Central to the issue of fragmentation in a self-gravitating accretion
disk is the cooling rate of the material.  If the disk is formally
unstable, gravitationally bound clumps appear only if the cooling time
is shorter than the orbital period about the central object
\citep[e.g.,][]{gam01}; these points are reviewed in
\S~\ref{sec:disk}.  Radiation leakage is far too slow in collapsar
disks to be an effective coolant \citep[e.g.,][]{npk01}.  However,
such disks have sufficiently high temperatures ($\ga$1\,MeV) and
densities ($\ga$$10^8\g\cm^{-3}$) that nuclear processes play critical
roles in the energy budget.  We argue in \S~\ref{sec:cool} that
nuclear photodisintegration is the only energy sink fast enough to
permit fragmentation.  The condition for gravitational instability and
the strong temperature sensitivity of the photodisintegration rate
allow us to nicely constrain the location, size, and mass of an
unstable clump.  Our estimates suggest that a fragment will collapse
until it is supported by neutron degeneracy pressure, forming a
low-mass ($\approx0.1-1\msun$) neutron star.

As described in \S~\ref{sec:migration}, a fragment will be driven
toward the black hole as disk stresses and gravitational radiation
remove orbital angular momentum.  Migration ends with the tidal
disruption of the fragment.  We calculate the time scales for
migration and subsequent accretion, and consider whether fragmentation
accounts for the X-ray flares in long GRBs.  Assuming that only a
single fragment orbits the central black hole, we estimate in
\S~\ref{sec:ligo} the characteristic gravitational-wave strain from
this binary using a total migration rate that includes viscous
torques. 


\section{Gravitational Instability in Collapsar Disks}
\label{sec:disk}

Here we consider the conditions for gravitational instability and
fragmentation of {\em steady-state} collapsar disks \citep[see
also][]{npk01,dpn02}, where the mass inflow rate is constant in
radius. The steady-state assumption is a gross simplification,
especially at large disk radii \citep{mw99}, where material in the
collapsing star with high specific angular momentum may continue to
rain in for many viscous times.  Since this region of the disk is the
staging ground for our study, our results should be viewed as only
rough approximations.

Our background model consists of a central black hole of mass $M_{\rm
BH}$ surrounded by a Keplerian accretion disk with radially varying
orbital frequency $\Omega = (GM_{\rm BH}/r^3)^{1/2}$, surface density
$\Sigma$, isothermal sound speed $c_s$, and vertical scale height $H =
c_s/\Omega$.  Mass flows at a rate of $\dot{M} = 3\pi\nu\Sigma$
\citep[e.g.,][]{pri81}, where $\nu = \alpha c_s H$ is the usual
viscosity prescription \citep{ss73}.  We assume that $H/r = \eta$ is a
fixed parameter, which is physically plausible in the thick, advective
outer region of a steady collapsar disk \citep[e.g.,][]{pwf99}.
Results below are given in term of the scaled variables $r_{100}
\equiv r/(100GM_{\rm BH}c^{-2})$, $M_3 \equiv M_{\rm BH}/3\msun$,
$\mdot\equiv\dot{M}/1\,M_\odot\,{\rm s^{-1}}$, $\al \equiv
\alpha/0.1$, and $\de \equiv \eta/0.5$.

Gravitational instability arises when \citep{too64,glb65}
\be\label{eq:toomre}
Q\equiv\frac{\Omega c_s}{\pi G\Sigma} < \qc \simeq 1~.
\ee
The value of $\qc$ depends on the disk thickness and the equation of
state.  The disk equations above imply that $Q\propto
\alpha\eta^3\dot{M}^{-1} r^{-3/2}$ decreases with $r$, and thus
instability holds for
\be\label{eq:radius}
r_{100} \ga 4 \al^{2/3}\de^2\mdot^{-2/3}~.
\ee
When $Q < Q_{\rm crit}$, the linearly unstable mode with the fastest
growth rate ($\approx$$\Omega/Q$) has a wavelength of $\approx$$QH$
and mass of (e.g., \S~6.3 of Binney \& Tremaine 1987)
\be\label{eq:m_frag}
(QH)^2\Sigma \approx 0.02 Q^2 
 \al^{-1} M_3 \mdot r_{100}^{3/2} \msun~.
\ee
We identify $(QH)^2\Sigma$ with the mass of a bound clump if cooling
is rapid enough to permit collapse (see \S~\ref{sec:cool}).

Numerical simulations of thin disks with idealized equations of state
and cooling prescriptions show that gravitational instability leads to
fragmentation when the cooling time scale satisfies $\Omega t_{\rm
cool} < \xi$, where $\Omega \approx 70\,M_3^{-1}
r_{100}^{-3/2}\s^{-1}$, and $\xi \approx 1$--10
\citep[e.g.,][]{gam01,mej05,ric05}.  When cooling is slow,
gravitational instability produces spiral waves that dissipate as
thermal energy, driving the disk back toward stability.  It is unclear
if we can reliably use previous simulations to draw conclusions about
collapsar disks, but we take these results as a plausible starting
point. 


\section{Cooling Mechanisms}
\label{sec:cool}

Using the disk equations in \S~\ref{sec:disk}, we evaluate the
discriminant $\Omega\tc=\Omega\Sigma c_s^2/qH$ for various possible
cooling mechanisms, where $q$ is the relevant cooling rate in units of
${\rm erg\ s^{-1}\ cm^{-3}}$. Cooling via radiative diffusion can be
ignored outright, since huge optical depths lead to photon diffusion
times of many years \citep{npk01}.

The most important neutrino cooling processes are $e^\pm$ pair
annihilation to neutrinos and the Urca process. We adopt the
corresponding cooling rates in \citet{pwf99}. These rates are strong
functions of disk temperature, which for an ideal gas is $T_{10}\equiv
T/10^{10}\,{\rm K} \approx 3.6 \mu \de^2 M_3 r_{100}^{-1}\,{\rm K}$,
where $\mu$ is the mean molecular weight.  Here we assume for
calculational purposes that material at $r_{100} > 1$ is pure helium
($\mu = 4/3$) and has the ideal-gas equation of state.  At the radius
where $Q=Q_{\rm crit}$ (see eq. [\ref{eq:radius}]), cooling from pair
annihilation gives
\be 
	(\Omega\tc)_{\rm pair} 
   \approx 800\al^{7/3}\de^{-9}M_3^{-3}\dot{M}_1^{-7/3}~,
	\label{eq:pair}
\ee 
A similar calculation for Urca cooling yields
\be 
(\Omega\tc)_{\rm Urca} \approx
30\al^{7/3}\de^{-3}M_3^{-1}\dot{M}_1^{-7/3}
X_{\rm nuc}^{-1}~,
	\label{eq:urca}
\ee
where $X_{\rm nuc}$, the mass fraction of free nucleons, is very
sensitive to $T$ \citep[e.g.,][]{pwf99}.  Both $(\Omega\tc)_{\rm
pair}$ and $(\Omega\tc)_{\rm Urca}$ seem too large to allow
fragmentation when $\al \approx 1$ \citep[a natural choice for
self-gravitating disks;][]{gam01,ric05} and $X_{\rm nuc} \approx 1$.
While we can not rule out pair annihilation as a relevant coolant (for
instance, when $\al \approx 0.1$), Urca cooling is important only
where nuclear photodisintegration is nearly complete.  As we now
demonstrate, photodisintegration itself will absorb sufficient energy
quickly enough to allow fragmentation.

For photodisintegration of $^4{\rm He}$ we set
\be
(\Omega \tc)_{\rm photo} = \frac{\Omega c_s^2}{E\lambda}
\approx 20 \de^2 M_3^{-1} r_{100}^{-5/2} \lambda^{-1}~,
\ee
where $E=7.07\ {\rm MeV}/m_p$ is the $^4{\rm He}$ binding energy per
nucleon and $\lambda$ is the reaction rate in ${\rm s}^{-1}$.
Experiments show that the $^4{\rm He}(\gamma,{\rm n})^3{\rm He}$ and
$^4{\rm He}(\gamma,{\rm p}){\rm T}$ cross sections rise abruptly at
photon energies of $\approx$$20\ {\rm MeV}$ to a plateau value of
$\sigma\approx$$1\ {\rm mb}$ \citep[$10^{-27}\,{\rm cm}^2$;][]{fel90}.
Subsequent photodisintegration steps have larger cross sections and
smaller threshold energies \citep{sko81,bir85}, contributing little to
the overall rate.  At temperatures $kT \la 20\,{\rm MeV}$ ($T_{10} \la
23.2$), we approximate the $^4{\rm He}$ photodisintegration cross
section as a step function and integrate over the Wien tail of the
Planck spectrum to estimate
\be 
    \lambda &\approx& 10^{17} T_{10}
    \,\exp\lp-\frac{23.2}{T_{10}}\rp{\rm s^{-1}}~,
    \label{eq:rate}
\ee
which is a good fit to detailed calculations provided to us by
H. Schatz (2006, private communication).  Because of the exponential
factor, $\lambda$ varies from $\approx$1 to $\approx$$10^{9}$ as
$T_{10}$ is increased from 0.6 to 1.2.  Thus, photodisintegration
turns on sharply at $r_{100} \approx 6\de^2$, interior to which
$(\Omega \tc)_{\rm photo}$ plummets to $\ll$1.  Rapid
photodisintegration may initiate at slightly larger radii if a
gravitationally unstable clump has some compressional heating as it
begins to collapse.

It seems that photodisintegration is a very effective coolant,
permitting fragmentation over a small range of radii near the location
where $Q = Q_{\rm crit}$ in the unperturbed disk.  By substituting the
radius at which $Q \simeq 1$ (see eq.~[\ref{eq:radius}]) into
eq.~(\ref{eq:m_frag}), we estimate the mass of a bound fragment:
\be\label{eq:m_frag2}
	M_f \approx 0.2 \de^3 M_3 \msun.
\ee
Multiple fragments may form in the same region \citep[e.g.,][]{ric05},
although it is unclear how their individual masses are distributed.
We assume in \S\S~\ref{sec:migration} and \ref{sec:ligo} that small
fragments merge into a single body of mass $\approx$0.1--$1\msun$.

\begin{figure}
  \centerline{\epsfig{file = 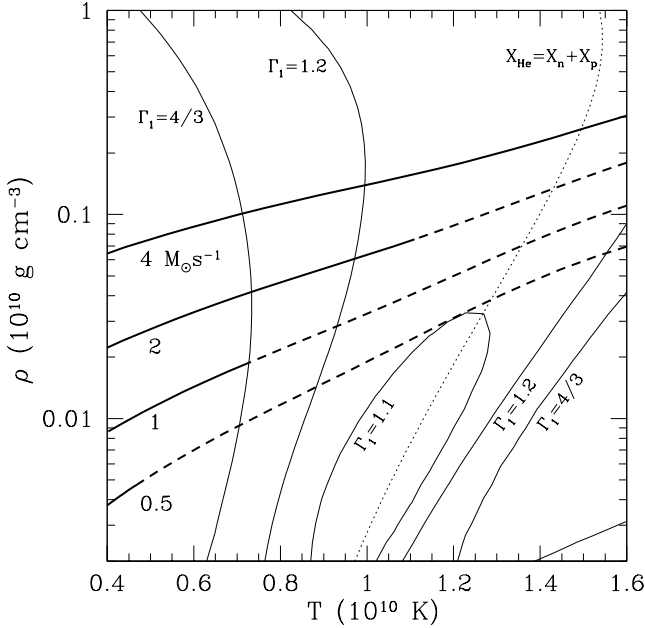, angle = 0, width = 1.05\linewidth}}
  \caption{Properties of a plasma composed of $^4{\rm He}$, protons,
    neutrons, electrons, and positrons as a function of density and
    temperature. The thin solid lines are contours of constant
    $\Gamma_1$, and the thin dotted line marks the half
    photodisintegration contour ($X_{\rm He}= X_n+X_p$). Thick lines
    represent disk models with $\alpha=0.1$, $\eta=0.5$, and the
    accretion rates shown.  Dashed and solid portions of these curves
    indicate whether the disk is gravitationally stable or unstable
    ($Q<Q_{\rm crit}$), respectively.
	 \label{fig:gamma}\vspace{2mm}}
\end{figure}

The breakdown of $^4{\rm He}$ into free neutrons and protons has two
added benefits.  First, efficient Urca cooling becomes possible,
promoting the continued collapse of the fragment.  Second, as
photodisintegration proceeds and $X_{\rm nuc}$ approaches unity, the
adiabatic pressure-density exponent $\Gamma_1\equiv(\partial\log
P/\partial\log \rho)_{\rm ad}$ may dip below 4/3, in which case the
gravitationally bound fragment is unstable and collapses dynamically.
This second point is illustrated by \citet{in66}, who compute the
thermodynamic properties of a plasma composed of $^{56}{\rm Fe}$,
$^4{\rm He}$, neutrons, protons, electrons, and positrons.  We have
repeated these calculations\footnote{The assumption of thermodynamic
equilibrium is not strictly valid unless the material is opaque to
neutrinos \citep{bel03}, but the approach of \citet{in66} is still
useful as an illustrative tool.} without $^{56}{\rm Fe}$.
Figure~\ref{fig:gamma} shows the results in the temperature-density
plane, where we see that the trajectories of our disk models pass
through the region with $\Gamma_1 < 4/3$.  However, the overlap of $Q
< Q_{\rm crit}$ (thick, solid curves) and $\Gamma_1 < 4/3$ occurs only
for sufficiently high $\dot{M}$ (e.g., $\ga$$1\msun\,{\rm s}^{-1}$).

Collapse of the fragment halts once the density is high enough for
neutron degeneracy to provide pressure support.  At this point we
might expect the fragment's radius to be
$R_f \approx 10 (M_f/M_\odot)^{-1/3}\,{\rm km}$.
However, there are numerous uncertainties regarding the formation and
ultimate structure of this newly formed low-mass neutron star,
including the hydrodynamics of the collapse, the equation of state,
the effects of the neutrino pressure and finite temperature, and the
fate of the fragment if $M_f \la 0.1\msun$, the minimum mass of a
stable neutron star \citep[e.g.,][]{cst89}.  An accurate, quantitative
assessment of these uncertainties is not possible at present.

%


\section{Migration and Tidal Disruption}
\label{sec:migration}

A net torque on the orbit of the fragment arises from dissipation
within the disk and the loss of orbital angular momentum in the form
of gravitational waves.  When the fragment is massive enough to open a
gap in the disk, it migrates inward on the viscous time scale
$r/\dot{r} \approx r^2/\nu$ \citep[i.e., type II migration;][]{lp86},
or
\be\label{eq:tvisc}
t_\nu \approx (\alpha\eta^2\Omega)^{-1} 
\approx 1\,\al^{-1}\de^{-2} M_3 r_{100}^{3/2}\,{\rm s}~.
\ee
Gap formation is expected when $M_f/M_{\rm BH} \ga \alpha^{1/2} \eta^2
\approx 0.08\,\al^{1/2}\de^2$ \citep[e.g.,][]{tak96}, a result derived
in relation to planet formation.  For simplicity, we adopt $t_\nu$ as
the migration time from disk torques, even if the gap condition is not
strictly satisfied.  Loss of orbital angular momentum in gravitational
waves causes inspiral on a time \citep{pet64}
\be\label{eq:tgw}
	t_{\rm gw} =
	\frac{5}{64\Omega}\lp\frac{G\calm}{c^3}\Omega\rp^{-5/3}
	\approx 700\,\calm_1^{-5/3} M_3^{8/3} r_{100}^4\,{\rm s}
	~,
\ee
where $\calm \equiv \calm_1 M_\odot \approx M_f^{3/5}M_{\rm BH}^{2/5}$
is the chirp mass.  The two migration times are equal when the orbital
frequency is
\be\label{eq:omeq}
	\Omega_{\rm eq} \approx 
	\frac{c^3}{G\mathcal{M}}
	\lp\frac{5\alpha\eta^2}{64}\rp^{3/5}
	\simeq 4.8\,\al^{3/5}\de^{6/5} \calm_1^{-1}\, {\rm kHz}~.
\ee
For $\Omega > \Omega_{\rm eq}$, gravitational radiation yields the
dominant torque.  Both $\Omega t_\nu \gg 1$ and $\Omega t_{\rm gw}\gg
1$ when $r \gg GM_{\rm BH}/c^2$, so that an initially circular orbit
remains nearly circular during migration.  Also note that efficient
neutrino cooling at $r_{100} \la 1$ allows the disk to thin
substantially \citep[$\eta \approx 0.1$--0.3;][]{pwf99}, increasing
$t_\nu$ and decreasing $\Omega_{\rm eq}$.  This is important in
\S~\ref{sec:ligo}, where we estimate the gravitational-wave signal.

Tidal disruption of the fragment begins when it fills its Roche lobe
of radius $\approx$$0.5r(M_f/M_{\rm BH})^{1/3}$.  If the fragment has
the radius of a neutron-degenerate object (see \S~\ref{sec:cool})
disruption occurs at frequency
\be
\Omega_{\rm dis} \approx 3.5\,(M_f/M_\odot)\,{\rm kHz}~.
\ee
Whether the fragment is disrupted dynamically or undergoes stable mass
transfer driven by gravitational radiation \citep[e.g.,][]{bc92}, most
of the fragment's mass is stripped in $<$1\,s.  

Given the short time for disruption, it seems unlikely that
fragmentation of the sort discussed here and in \citet{paz06} can
account for the late-time X-ray flares in long GRBs, which have rise
and decay times similar to their onset times of $\ga$$10^2\,{\rm
s}$. Moreover, if a fragment forms at $r_{100} \approx 1$--4 and
viscous stresses set the migration time to $\approx$10--100\,s,
accretion of the fragment might blend with the prompt emission.
The migration timescale may be increased if the fragment is especially massive in
comparison to the disk \citep{cs96}, but probably by a factor of $<10$.
Better understanding of the interaction between the fragment and the
disk will require detailed hydrodynamical simulations.  Gravitational
instability at large disk radii may power the GRB flares, but, if so,
probably through enhanced angular momentum transport rather than
fragmentation.


\section{Detection of Gravitational Waves}
\label{sec:ligo}

LIGO and similar detectors, such as VIRGO, GEO600, and TAMA300, are
sensitive to gravitational waves with frequencies of
$\sim$10--$10^3\,{\rm Hz}$.  For the full range of GRB models,
gravitational waves in the LIGO band would be generated within the
central engine, either the compact binary progenitor of a short burst
or in a collapsar disk.  Measurement of the gravitational-wave signal
from a GRB would be a unique probe of the physics at the heart of the
explosion \citep[see][]{kp93,fin99,vp01,km03}.  Previous work on
gravitational radiation from collapsar disks have addressed general
nonaxisymmetric instabilities in compact accretion tori \citep{vp01}
or bar-mode instabilities \citep{km03}.  Unlike these earlier
scenarios, the simplest form of our fragmentation model constrains the
mass quadrupole moment and its evolution, features we now exploit to
obtain the gravitational-wave signal.

As in the previous section, we imagine a binary composed of a single
fragment in a circular orbit around the black hole.  The measured
gravitational-wave strain is $h(t) = h_0(t) \cos\phi(t)$, where $f
\equiv 100\,f_{100}\,{\rm Hz} =\Omega/\pi$ is the slowly varying wave
frequency, and $\phi = 2\pi \int dt f$ is the accumulated phase.  For
a source at distance $D = 100\,D_{100}\,{\rm Mpc}$, the strain
amplitude is
\be\label{eq:strain}
h_0  & = & \Theta \frac{G\calm}{c^2D}
\left(\frac{\pi G\calm }{c^3}f\right)^{2/3} \nonumber \\
& \simeq &
6.4\times 10^{-24}\Theta \calm_1^{5/3} f_{100}^{2/3} D_{100}^{-1}~,
\ee
where $\Theta = 0$--4 is a factor incorporating the orientation of the
source and the antenna pattern of the detector \citep[e.g.,][]{fc93}.
For completely random source orientations and positions on the sky,
the root-mean-square of $\Theta$ is $\sqrt{\langle \Theta^2\rangle} =
1.6$.  Long GRBs seem to be narrowly beamed, probably in a direction
normal to the accretion disk.  Therefore, we expect an associated
black-hole/fragment binary to have a low inclination, which gives
$\sqrt{\langle \Theta^2\rangle} \simeq 2.5$, or $\simeq$60\% larger
than the random case \citep[see also][]{kp93}.  In what follows, we
let $\Theta = 2.5$.

When the waveform is known, the matched filter approach can be used to
search for the signal.  The signal-to-noise for a slow binary inspiral
is then \citep[e.g.,][]{fh98}
\be\label{eq:snr}
\lp\frac{S}{N}\rp^2 = \int d(\ln f) \, n_{\rm cyc}(f) \,
\frac{h_0^2(f)}{h_n^2(f)}~,
\ee
where $n_{\rm cyc} \equiv f^2/\dot{f}$ is the number of cycles per
unit $\ln f$, and $h_n$ is the strain noise in a bandwidth
$\approx$$f$ centered on $f$. We see that $\sqrt{n_{\rm cyc}}$ is the
signal gain and $\sqrt{n_{\rm cyc}}\,h_0\equiv h_c$ is the associated
characteristic strain.

\begin{figure}
  \centerline{\epsfig{file = 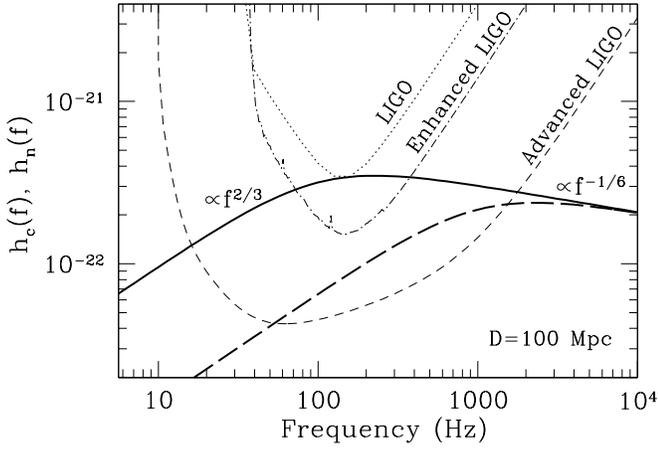, angle = 0, width = 1.05\linewidth}}
  \caption{Characteristic gravitational-wave strain $h_c(f)$ from the
  	inspiraling black-hole/fragment binary for $f_{\rm eq}=100\ {\rm
  	Hz}$ ({\it thick solid line}) and $1\ {\rm kHz}$ ({\it thick dashed
  	line}).  Both curves assume $\calm=1\msun$, $D=100\ {\rm Mpc}$, and
  	$\Theta=2.5$.  The thin lines show the broadband strain noise for
  	current LIGO ({\it dotted}), enhanced LIGO\protect\footnotemark[2] ({\it dot-dashed}), and
	advanced LIGO ({\it dashed}).
	\label{fig:strain}}
\end{figure}

\begin{samepage}
\footnotetext[2]{For further information see "Enhanced LIGO", LIGO internal document
T060156-01-I, R. Adhikari, P. Fritschel, and S. Waldman available at
http://www.ligo.caltech.edu/docs/T/T060156-01.pdf}
\end{samepage}

If $|\dot{r}/r| = t_\nu^{-1} + t_{\rm gw}^{-1}$ is the total inward
migration rate from \S~\ref{sec:migration}, then $n_{\rm cyc} =
2fr/3\dot{r}$. Some minor algebra yields
\be
n_{\rm cyc}(f) 
= \frac{2}{3} f t_{\rm gw}
\left[
  1 + \lp\frac{f_{\rm eq}}{f}\rp^{5/3}
  \right]^{-1}~,
\ee
where $f_{\rm eq} = \Omega_{\rm eq}/\pi$ (see eq.~[\ref{eq:omeq}])
tunes the importance of viscous torques.  When $f \ll f_{\rm eq}$
viscous migration dominates and $h_c \propto f^{2/3}$, while
gravitational-wave driven inspiral ($f \gg f_{\rm eq}$) gives $h_c
\propto f^{-1/6}$.  Figure~\ref{fig:strain} shows $h_c(f)$ for $f_{\rm
eq} = 100\,{\rm Hz}$ and 1\,kHz.  The lower value of $f_{\rm eq}$
essentially assumes that the disk is fairly thin ($\eta\approx0.1$),
as expected when neutrino cooling is efficient. Since $h_c$ is peaked
near $f_{\rm eq}$, albeit rather broadly, we can use the approximation
$S/N\approx h_c(f_{\rm eq})/h_n(f_{\rm eq})$.  Inspection of
Figure~\ref{fig:strain} indicates that, for $D_{100} \sim 1$, $S/N
\approx 5$--10 when $f_{\rm eq} = 100\,{\rm Hz}$, while $S/N \la 1$
when $f_{\rm eq} = 1\,{\rm kHz}$.  Detection with advanced LIGO seems
quite possible for nearby GRBs. The combined effects of viscous
torques and gravitational-wave emission on the orbital evolution
create a distinct strain signal, where an estimate of $f_{\rm eq}$
would provide unique insight into the physics of collapsar disks.

Estimates of the local rate of detectable long GRBs span
$\approx$0.1--$1\,{\rm Gpc}^{-3}\,{\rm yr}^{-1}$
\citep[e.g.,][]{gpw05}, based largely on a small sample with moderate
to high redshifts.  This implies a rate of $\la$$4\times 10^{-3}
D_{100}^3\,{\rm yr}^{-1}$ within a distance $D$ .  However, two GRBs
in the past 8 years within 150\,Mpc \citep[980425 and 060218;
e.g.,][]{gal98,fer06} suggest tentatively that the local rate is more
uncertain and could be as large $\approx$$0.1\,D_{100}^3\,{\rm
yr}^{-1}$, which is consistent with $10^{-4}$ of the local rate of core-collapse
supernovae \citep{cap99}.
Beaming implies there may be
$\approx$100 unseen GRBs for every detectable event, giving a total local
GRB rate of $\la$$10\,D_{100}^3\,{\rm yr}^{-1}$. If a misdirected GRB is
detected by multiple gravitational-wave antennae, it may be possible to locate
the source to within a few degrees and provide a target region to search for an
``orphaned'' electromagnetic afterglow.



\acknowledgments 

We are grateful to Lars Bildsten for critical comments, Scott Hughes
for suggestions on a previous draft, and Hendrik Schatz
for sharing photodisintegration rates. We thank Rana Adhikari and David Shoemaker
for providing LIGO noise strain estimates.
We also thank Andrei Beloborodov, Wen-Xin Chen, and William Lee
for answering questions about collapsar disks.
This work was
supported by NSF grant PHY 99-07949 and by the Joint Institute for
Nuclear Astrophysics through NSF grant PHY 02-16783.


\end{document}